# Deep learning-based synthetic CT generation from MR images: comparison of generative adversarial and residual neural networks

Faeze Gholamiankhah[1], Samaneh Mostafapour[2], and Hossein Arabi[3]

[1]Department of Medical Physics, Faculty of Medicine, Shahid Sadoughi University of Medical Sciences, Yazd, Iran.
[2]Department of Radiology Technology, Faculty of Paramedical Sciences, Mashhad University of Medical Sciences, Mashhad, Iran.
[3]Division of Nuclear Medicine and Molecular Imaging, Department of Medical Imaging, Geneva University Hospital, CH-1211 Geneva 4, Switzerland.

**Abstract**

**Background:**

Currently, MRI-only radiotherapy (RT) eliminates some of the concerns about using CT images in RT chains such as the registration of MR images to a separate CT, extra dose delivery, and the additional cost of repeated imaging. However, one remaining challenge is that the signal intensities of MRI are not related to the attenuation coefficient of the biological tissue. This work compares the performance of two state-of-the-art deep learning models; a generative adversarial network (GAN) and a residual network (ResNet) for synthetic CTs (sCT) generation from MR images.

**Materials and methods:**

The brain MR and CT images of 86 participants were analyzed. GAN and ResNet models were implemented for the generation of synthetic CTs from the 3D T1-weighted MR images using a six-fold cross-validation scheme. The resulting sCTs were compared, considering the CT images as a reference using standard metrics such as the mean absolute error (MAE), peak signal-to-noise-ratio (PSNR) and the structural similarity index (SSIM).

**Results:**

Overall, the ResNet model exhibited higher accuracy in relation to the delineation of brain tissues. The ResNet model estimated the CT values for the entire head region with an MAE of 114.1±27.5 HU compared to MAE=-10.9±147.0 HU obtained from the GAN model. Moreover, both models offered comparable SSIM and PSNR values, although the ResNet method exhibited a slightly superior performance over the GAN method.

**Conclusion:**

We compared two state-of-the-art deep learning models for the task of MR-based sCT generation. The ResNet model exhibited superior results, thus demonstrating its potential to be used for the challenge of synthetic CT generation in PET/MR AC and MR-only RT planning.

## 1. Introduction

Computed tomography (CT) plays a significant role in treatment planning and dose calculation in the radiation therapy (RT) chain by providing 3-dimensional attenuation coefficient maps. These are used to calculate organ and tissue-specific doses (1). Modern techniques, such as intensity-modulated radiation therapy (IMRT) and volumetric-modulated radiation therapy (VMAT), rely on anatomical images to accurately define the target and organs at risk (OAR) for proper dose delivery (1, 2). In clinical practice, the use of magnetic resonance imaging (MRI) for treatment planning is increasing due to the high contrast soft-tissue discrimination and sharper organ boundaries possible in comparison with CT imaging. Moreover, some studies have shown that functional MRI information, including diffusion-weighted imaging (DWI) and dynamic contrast-enhanced imaging, could aid in identifying active tumor sub-volumes in head and neck cancer (3). Currently, MR images are integrated into the RT chain through a rigid or deformable registration to the reference CT image for the precise delineation of the target volume and OAR. The electron density information from CT images is used for dose calculations (2, 3). However, errors associated with MR to CT image registrations introduce a systematic uncertainty leading to a significant dosimetric impact, particularly for small tumors in the vicinity of OARs (4, 5). To avoid these errors in RT planning as well as to reduce the cost of therapy, MRI-only RT planning is introduced which only relies on MR images in the radiotherapy workflow. MRI-only RT obviates the need for MR and CT image registration and additionally decreases the number of imaging sessions (CT imaging) and its associated costs in a workflow. This leads to a reduction in the received dose, particularly for the patients requiring multiple scans during their treatment process (1-3, 5). However, MRI-only RT planning faces the challenge of geometric distortion due to the magnetic field non-uniformity, the absence of a cortical bone signal in conventional MR images (6, 7), and the lack of an attenuation coefficient map (8, 9). The primary challenge for MR-only RT planning stems from the fact that the signal intensities of MRI correlate with the tissue proton density and tissue relaxation properties. MR signals do not relate to the photon attenuation coefficients of the tissues. On the other hand, the voxel intensity of the CT images directly reflects the radiological characteristics of the tissue (1). The same challenge is faced by the MR-based attenuation correction (MRAC) in the hybrid PET/MR to convert the patient's MR image into an attenuation coefficient map. In this regard, a number of approaches have been proposed for the generation of synthetic (pseudo) CT images from MRI data (10-12).

There are three major categories of methods for synthetic CT generation: tissue segmentation (13), atlas (6, 8), and artificial intelligence (14). Tissue segmentation-based approaches create attenuation or photon coefficient maps via the bulk segmentation of MR images into few tissue classes followed by the assignment of their corresponding coefficient values (13, 15). Discriminating between bone and air tissues is one of the major challenges since bone and air have very low and roughly similar signals on conventional MR sequences. The use of other MR sequences, such as the ultra-short echo time (UTE) and zero-echo-time, has eliminated this problem. However, these sequences also suffer from an increased scanning time or low signal-to-noise ratio (15, 16). The atlas-based approach consist of deformable registration algorithms for the purpose of aligning the target MRI to the numbers of MRIs in an atlas database, followed by the assignment of CT numbers in the atlas database for each voxel of the target MRI (11, 17).

Recently, machine learning, especially convolutional neural networks (CNN), has emerged as a promising approach to improve the quality of medical image analysis including image segmentation, denoising, reconstruction, and particularly synthesizing pseudo-CTs from MR images (18, 19). Many studies have been conducted to address the challenge of synthetic-CT generation from MR images using different algorithms/architectures or convolutional neural networks. However, only a few deep learning models are being frequently used due to their robust, accurate, and reliable performance. Generative adversarial networks (GAN) and residual networks are among the highly popular deep learning models that have shown promising results in various fields of medical image analysis (20-25). GAN networks, owing to their sophisticated architecture benefiting from the generator and discriminator compartment, and residual networks, owning to their large receptive fields, are able to offer relatively optimal solutions for a vast range of image-related problems such as transformation and segmentation.

This study set out to compare two state-of-the-art deep learning models, specifically the generative adversarial network (GAN) and residual network, for the task of MR-guided synthetic CT generation. Although many approaches/algorithms have been proposed in the previous works concerning the generation of synthetic (pseudo) CT images from MRI data (6-8, 13, 14), deep learning-based approaches are of special interest owing to their promising and superior performance (10, 26). Among the various deep learning models, GAN and residual deep learning models are frequently used for different purposes in clinical and research settings (25-29). The major aim of this study was to compare the two popular deep learning models for the challenging task of MR-guided synthetic CT generation related to their application in MR-only radiation planning (11, 19) and MR-guided PET attenuation correction (11).

## 2. Materials and Methods

### 2.1 CT and MRI data acquisition

The patient population consisted of 46 men (mean age: 61±12 years, mean weight: 79.3±11 kg) and 40 women (mean age: 57±7 years, mean weight: 71.2±10 kg) who underwent brain CT and MRI scans. The clinical indications included neurodegenerative disease (40 men and 30 women), epilepsy (3 men and 5 women), and different graded brain tumors (3 men and 5 women). This study was approved by the Ethics code of 241345CH (date: 20/10/2018). The MRI scans were performed using 3T MAGNETOM Skyra (Siemens Healthcare, Erlangen, Germany) with a 64-channel head coil using a T1-weighted (magnetization-prepared rapid gradient-echo (MP-RAGE)) sequence and the parameters of TE/TR/TI, 2.3 ms/1900 ms/970 ms, flip angle 8º; NEX = 1. The T1-weighted MR images were saved in a matrix dimension of 255×255×250 with a voxel size of 0.86×0.86×1 mm. The CT image acquisitions with 120 kVp and 20 mAs were performed on a Toshiba Aquilion (Toshiba Co., Tokyo, Japan). The matrices of CT images were 512×512×149 voxels with a voxel size of 0.97×0.97×1.5 mm.

Due to the fact that the MRI and CT image acquisitions were not performed simultaneously, the MR images were aligned to the corresponding CT images. To this end, a mutual information-based image registration algorithm, which performs a combination of rigid and non-rigid deformation implemented in Elastix platform (based on the ITK library) (https://elastix.lumc.nl/, Netherlands), was employed to align the MR and CT images. Afterwards, the resolution of the aligned MR images was converted to the resolution of the corresponding CT images as a preprocessing step for the training and validation processes.

### 2.2 Network architecture

This study set out to compare two state-of-the-art deep convolutional neural network algorithms in the context of MR-guided synthetic CT generation. These include the ResNet and GAN models which have been extensively employed for the task of image segmentation and inter-modality image regression. In the following sections, the architecture of the two deep learning models has been described.

**2.2.1 ResNet architecture**

Deep residual networks, formed by a number of residual blocks, were introduced by He *et al.* to address the degradation problem in the training process of deep neural networks and to reduce the computational cost (30). Residual or shortcut connections result in skipping one or more layers in a network to address the gradient vanishing issue causing the direct propagation of signals in forward and backward paths from one block to other blocks (Figure 1). It should be noted that a network with *n* residual blocks which has $2^n$ unique paths would result in decreasing the effective receptive field. Therefore, the incorporation of residual connections in the training of a network would reduce the border effects of convolution leading to decreased distortion near to the borders.

The proposed architecture of ResNet, illustrated in Figure 2, consists of 20 convolutional layers wherein every two convolutional layers are stacked together by residual connections. Each convolutional layer is composed of an element-wise rectified linear unit (ReLU) and a batch normalization (BN) layer. The network takes the MR images as the input and provides stimulated CT images as the output. In the initial layers, 3×3×3 filters are applied that are related to the low-level image features. To extract the mid-level and high-level image features, the number of kernels is multiplied by a factor of two or four in the deeper layers. The output of the final layer, the fully connected softmax layer, is in the same dimension as that of the input image (20).

**2.2.2    GAN architecture**

General adversarial networks (GANs) were suggested by Goodfellow et al. in 2014. This model type consists of two adversarial generative and discriminative components that are trained simultaneously. The generator model learns to generate new data while the discriminator determines the probability of whether the input is data generated by the generator (fake) or real. The usage of adversarial nets when both models are multilayer perceptrons is more straightforward. In this regard, the samples are generated by passing random noise through a multilayer perceptron generator with a differentiable function that represents a mapping to the data space with a parameter of $\theta\_G$. The second multilayer perceptron is the discriminator with a parameter of $\theta\_D$ which determines the likelihood of false or true for the input data.

The optimization of adversarial nets is similar to the optimization of a two-player zero-sum minimax game conducted by jointly optimizing the cost functions of the discriminator and generator. For this purpose, each of the $\theta\_D$ and $\theta\_G$ parameters were updated once in every iteration to decrease the values of the respective cost functions. As the discriminator is trained to enhance the differentiation ability, the generator is also trained to maximize the probability of the discriminator assigning a true label to the false (artificial) data. In other words, the generator is intended to generate data that has a minimum difference compared to real data (31).

In the generator network, first, random Gaussian noise that has zero mean and unit variance was projected followed by the ReLU activation function to form the first feature maps. In order to achieve the image with respective sizes, up-scaling layers were used. Each of the up-scaling layers is composed of a transposed convolution with 2×2 stride, and convolution with batch normalization (BN) and ReLU. They duplicate the size of the previous feature maps and halve the number of channels. The final layer has two parts: first a convolution with BN and ReLU, and second, a convolution with a hyperbolic tangent function without BN to maintain the true statistical features of the data.

The discriminator network accepts images and its correlated three-channel pixel coordinates as the input. The first convolutional layer with a kernel size of 5×5 and leaky ReLU (LReLU) as the activation function forms the initial feature maps that have the same size as the input image. By adopting Resnet blocks and down-scaling layers, the network duplicates the number of channels and halves the feature maps in each layer. Each Resnet layer has two convolutions, both with BN and LReLU, and each down-scaling layer has a 2×2 stride convolution with BN and LReLU. The final Logit includes a Resnet, a projection, an LReLU, and another projection respectively. Except for the first layer in the discriminator, all convolutional layers in the generator and discriminator employ a 3×3 kernel (32).

**2.3 Model implementation**

The training of the ResNet and GAN models was performed using 86 pairs of brain MR and CT images as the input/output respectively as part of a six-fold cross-validation scheme. To this end, these models were implemented in the NiftyNet platform (version 0.6.0, King's College London, UK) which is a publicly available pipeline for the realization of deep learning models. NiftyNet is built on TensorFlow which consists of common architectures and networks used for a deep learning approach which can be easily retrieved and optimized for different tasks. The application of the NiftyNet platform includes segmentation, regression, and image synthesis (33).

The training of the models was carried in a 2-dimensional setting wherein each pair of MR and CT trans-axial slices were considered to be a training sample. The following training parameters were set for both the ResNet and GAN models: batch size=30, sample per volume=2, learning rate =0.003-0.001, decay = 0.0001, optimizer=Adam, and loss function =L2.

During the training of the models, 5% of the training samples were dedicated to the evaluation of the models within the training to verify the risk of overfitting. The evaluation and training losses (errors) exhibited insignificant differences for both the ResNet and GAN models which show that there is no risk of overfitting. The training of the ResNet was completed in 15 epochs as the training loss reached its plateau while the training of the GAN model took 22 epochs to reach its optimal point.

**2.4 Evaluation strategy**

To evaluate the performance of the ResNet and GAN models, the resulting synthetic CT images were compared to the ground-truth CT images. In this regard, all CT images were segmented into major tissue types, including air, soft tissue, cortical bone, and total bone. The intensity thresholds of -450 HU, 150 HU, and 400 HU were applied for the segmentation of air, total bone, and cortical bone respectively. Voxels within the range of -450 to 150 HU were considered to be the soft-tissue mask. The assessment of the major anatomical structures extracted by the ResNet and GAN models was conducted using a dice similarity coefficient (Eq. 1) (15), relative volume difference (RVD) (Eq. 2), Jaccard coefficient (Eq. 3) (34) and sensitivity (S) (Eq. 4) over both the entire head and segmented regions:

$$DSC(A_r, A_S) = \frac{2 \mid A_r \cap A_s \mid}{\mid A_r \mid + \mid A_s \mid} \quad \text{Eq. 1}$$

$$RVD(A_r, A_S) = 100 \times \frac{\mid A_r \mid - \mid A_s \mid}{\mid A_s \mid} \quad \text{Eq. 2}$$

$$JC(A_r, A_S) = \frac{\mid A_r \cap A_s \mid}{\mid A_r \cup A_s \mid} \quad \text{Eq. 3}$$

$$S(A_r, A_S) = \frac{\mid A_r \cap A_s \mid}{\mid A_s \mid} \quad \text{Eq. 4}$$

Where $A_r$ and $A_S$ represent the intensities of the volume of interests in the reference CT images and synthetic CT images. Moreover, by considering the voxels within the above-mentioned regions, the mean error (ME) (Eq. 5), mean absolute error (MAE) (Eq. 6), root mean square error

(RMSE) (Eq. 7), and relative error (RE) (Eq. 8) metrics were computed in respect of the reference CT images. By assuming $dA(i) = (A_s(i) - A_r(i))$ wherein $A_s(i)$ and $A_r(i)$ stand for the i-th voxel intensity in the sCT and reference CT images, the formulae would be

$$ME = \frac{1}{N}\sum_{i=1}^{N} dA(i) \qquad \text{Eq. 5}$$

$$MAE = \frac{1}{N}\sum_{i=1}^{N} |dA(i)| \qquad \text{Eq. 6}$$

$$RMSE = \frac{1}{N}\sqrt{(dA(i))^2} \qquad \text{Eq. 7}$$

$$RE = \sum_{i=1}^{N} \frac{(A_s(i) - A_r(i))}{A_r(i)} \qquad \text{Eq. 8}$$

Where $N$ Indicates the number of voxels in the segmented region. Additionally, for the entire head, as a single volume of interest, peak signal-to-noise-ratio (PSNR) (Eq.9) and structural similarity index (SSIM), (Eq.10) quantifies the image quality that was calculated using the following Eqs:

$$PSNR = 10\log\left(\frac{I^2}{MSE}\right) \qquad \text{Eq. 9}$$

$$SSIM = \frac{(2\mu_r\mu_s + K_1)(2\delta_{rs} + K_2)}{(\mu_r^2 + \mu_s^2 + K_1)(\delta_r^2 + \delta_s^2 + K_2)} \qquad \text{Eq. 10}$$

In Eq. 9, $I$ denotes the maximum intensity value of the reference CT or synthetic CT images, and MSE denotes the mean square error. In Eq. 10, $\mu_r$ and $\mu_s$ are the mean intensity value, and $\delta_r$ and $\delta_s$ are the variance of the two corresponding CT images. Parameters $K_1 = (k_1 I)^2$ and $K_2 = (k_2 I)^2$ with the constants of $k_1 = 0.01$ and $k_2 = 0.02$ were defined to stabilize the division with small denominators.

## 3. Results

The training and evaluation of the GAN and ResNet models was carried out using a six-fold cross-validation scheme, thus the results reported in this section were calculated over the entire patient population.

Figure 4 shows the cross-sectional views of the generated synthetic CT images along with the corresponding MR and reference CT images. The visual investigation revealed that the sCT images generated by the ResNet model are less noisy and have a higher similarity to the real CT. Furthermore, the ResNet model outperforms the GAN model, leading to more accurate bone and air delineation.

Table 1 summarizes the mean and standard deviations of the ME, MSE, RMSE, RE, RVD, Dice, JC, Sensitivity, SSIM, and PSNR metrics computed over the sCT images resulting from the ResNet and GAN methods compared to the ground truth CT images for 86 subjects. The parameters were calculated within the air cavities, soft tissue, cortical bone, and total bone regions, as well as the entire head. On average, ME, MAE, and RMSE exhibited smaller errors for most parts of the head region through the ResNet method. This observation is in agreement with the CT value bias reflected in the RE and RVD parameters. Furthermore, both approaches offered comparable values for the Dice, JC, Sensitivity, SSIM, and PSNR metrics. Altogether, it can be observed that the ResNet method exhibited slightly superior accuracy over the GAN method. The boxplots of the quantitative metrics comparing the performance of the two methods by considering the real CT images as a reference, have been presented in Figure 5.

Figure 6 represents the axial views of the sCT and ground truth CT images together with the corresponding binary masks of soft tissue, total bone, cortical bone, and air cavities.

The quantitative accuracy of the CT value estimation using the two proposed methods was further assessed by the calculation of the Hounsfield unit differences and the relative error rate between the real CT and sCT images. It is evident from the results shown in Figure 7 that both approaches led to a comparable bias.

In addition to the region-wise analysis, a joint histogram analysis was conducted to display the voxel-wise correlation between the reference and estimated CT values. Figure 8 illustrates that the CT images generated by the GAN and ResNet methods are highly correlated with the reference

CT images. However, the correlation coefficient is slightly higher for ResNet (R2=0.98) than GAN (R2=0.97).

## 4. Discussion

MR-guided synthetic CT generation is an essential step in MR-only radiation planning and PET attenuation correction in relation to hybrid PET/MR scanners. Recent studies have demonstrated the promising performance of deep learning approaches to synthesize a "pseudo-CT" from MR-only images for the task of attenuation correction (35-38) as well as MR-only radiation planning (19, 39). These approaches have outperformed conventional synthetic CT generation approaches such as the atlas- and segmentation-based methods (14, 19, 40). In this light, a comparison of deep learning-based methods for challenging tasks (such as MR-guided PET attenuation correction and radiation dosimetry) is of particular interest in order to establish a robust framework with minimal errors. In this work, two state-of-the-art deep learning algorithms, namely the ResNet and GAN models, were evaluated for the estimation of the synthetic CT images from T1-weighted MRI images in relation to brain imaging. Their quantitative performance was assessed against the reference CT images. Though there are a number of deep learning architectures, the Resnet and GAN models are regarded as the most powerful, papular, and effective models owing to their exclusive properties/characteristics (41). The GAN architecture benefits from two generator and discriminator cores which enable the extraordinary capacity to capture/model the underlying structures/patterns in order to generate synthetic images. On the other hand, the Resnet architecture relies on a simpler structure. However, the entire image processing in the Resnet model is performed based on the full spatial resolution of the input image at the different layers which allows this model to estimate/predict the desirable outputs with outstanding accuracy and detail.

The GAN model used in this work has a residual architecture in the generator component which is similar to the model developed by Emami *et al.* (18) with a residual block for the generator and fully connected convolutional neural network (CNN) for the discriminator component. The GAN model proposed by Emami *et al.* (18) led to 89.3 ± 10.3 Hounsfield units (HU) for the mean absolute error (MAE) over the entire field of view for 15 brain scans, thus exhibiting superior performance over the CNN model along with an MAE of 102.4 ± 11.1 HU, although the overall

MAE of GAN and ResNet models in this study across the entire brain regions for 86 patients were 161.3±38.1 HU and 114.1±27.5 HU respectively. These results show a higher error rate compared to those obtained by Han *et al.* (39) (84.8±17.3 HU), Emami *et al.* (18) (89.30±10.25 HU), and Arabi *et al.* (25) (101 ± 40 HU). However, a comparison of these models based on the MAE would not be fair/reasonable as different patient populations were used in these studies. Han *et al.* (39) employed a CNN model trained by 18 subjects using a six-fold cross-validation procedure, Emami *et al.* (18) validated their model using a five-fold cross-validation framework for 15 patients and Arabi *et al.* (25) used 40 patients under a two-fold cross-validation scheme.

The other model assessed in this study was the ResNet model that benefits from dilated convolutional kernels that allow for the high-resolution processing of the input images at different layers or feature levels without increasing the complexity of the model. This architecture would be very effective for the regression processes wherein inter-modality image conversion is required with a high spatial resolution. Altogether, the ResNet model exhibited slightly superior performance over the GAN model. However, the GAN model could be implemented in a variety of architectures such as CycleGAN (42) which is able to show excellent performance in the specific tasks such as unsupervised learning. The ResNet model, owing to its high-resolution processing of the input images, might be a better option for end-to-end supervised image translation wherein specific anatomical features/structures are mapped/reflected in the resulting synthetic images. The high-resolution and end-to-end connections of the input and output images in the ResNet model allowed for effective synthetic CT generation from the T1 weighted MR images.

One of the limitations of this study is that it only focused on brain imaging. However, pelvis and thorax imaging are important as well in RT planning and PET AC. Synthetic CT generation from MR thorax images is highly challenging due to the presence of the lung and high heterogeneity of the tissues (8). Moreover, a comparison with other popular deep learning models such as U-net (providing a baseline to compare the other approaches) could add to the value of this work. Therefore, a comparison of different deep learning approaches should also be conducted in the thorax and pelvic regions to determine the most accurate and robust deep learning-based synthetic CT generation algorithm. Moreover, this study lacks an evaluation of the resulting synthetic CT images in terms of radiation dosimetry wherein the absorbed dose in the synthetic

CT images should be compared to the absorbed dose in the reference CT images to quantify the expected errors in MR-only RT planning.

## 5. Conclusion

The present study evaluated and compared two state-of-the-art and popular deep learning models frequently used in research settings for the challenging task of synthetic CT generation from MR images. It was demonstrated that the ResNet model (versus the GAN model) is able to generate accurate synthetic brain CTs from MR images for the task of MRI-only radiation therapy and attenuation correction in integrated PET/MRI scanners. However, the performance of these methods should also be evaluated in other body regions such as the pelvis and thorax.

# Tables

**Table 1.** Statistics of quantitative comparison between reference CTs and synthetic CTs generated by GAN and ResNet methods in terms of ME, MAE, RMSE, RE, RVD, Dice, JC, Sensitivity, SSIM, and PSNR. Results are averaged across 86 patients and reported in the form of average ± standard deviation.

| Res-Net method | air | Soft tissue | Cortical bone | Total Bone | Brain total |
|---|---|---|---|---|---|
| ME(HU) | 38.8±234.8 | 7.1±6.2 | -26.4±150.4 | -19.5±111.6 | -1.3±38.7 |
| MAE(HU) | 486.7±107.6 | 40.5±6.7 | 375.0±67.1 | 292.6±48.9 | 114.1±27.5 |
| RMSE(HU) | 153.9±9.6 | 57.1±5.2 | 157.4±7.0 | 156.0±6.0 | 91.5±5.8 |
| RE(%) | -0.43±0.09 | -0.20±0.07 | -0.11±0.17 | -0.02±0.18 | -0.09±0.16 |
| RVD(%) | 3.65±57.77 | 0.62±3.24 | -2.08±11.32 | -1.85±8.25 | -0.02±0.02 |
| Dice | 0.58±0.12 | 0.94±0.02 | 0.82±0.04 | 0.85±0.04 | 1.00±.0 |
| JC | 0.41±0.11 | 0.89±0.03 | 0.70±0.05 | 0.74±0.05 | 1.00±.0 |
| Sensitivity | 0.63±0.19 | 0.94±0.02 | 0.84±0.06 | 0.86±0.05 | 1.00±.0 |
| SSIM | - | - | - | - | 0.95±0.04 |
| PSNR | - | - | - | - | 28.65±1.59 |
| **GAN method** | **air** | **Soft tissue** | **Cortical bone** | **Total bone** | **Brain total** |
| ME(HU) | -213.2±125.0 | -12.1±29.2 | -5.9±159.9 | 34.0±115.6 | -5.4±57.5 |
| MAE(HU) | 501.0±60.1 | 67.2±18.5 | 443.6±94.7 | 323.1±57.4 | 161.3±38.1 |
| RMSE(HU) | 161.2±6.3 | 77.6±14.0 | 163.4±7.0 | 162.8±5.9 | 111.5±11.8 |
| RE(%) | -0.26±0.14 | -0.32±0.37 | -0.17±0.15 | -0.01±0.18 | 0.18±0.47 |
| RVD(%) | 59.38±34.26 | -12.21±11.32 | 6.40±16.37 | 35.02±41.57 | -0.10±0.26 |
| Dice | 0.53±0.07 | 0.87±0.07 | 0.75±0.07 | 0.73±0.10 | 1.00±.0 |
| JC | 0.37±0.06 | 0.78±0.10 | 0.61±0.08 | 0.58±0.12 | 1.00±.0 |
| Sensitivity | 0.44±0.08 | 0.94±0.02 | 0.74±0.09 | 0.66±0.14 | 1.00±.0 |
| SSIM | - | - | - | - | 0.94±0.05 |
| PSNR | - | - | - | - | 26.94±1.53 |

**Figures**

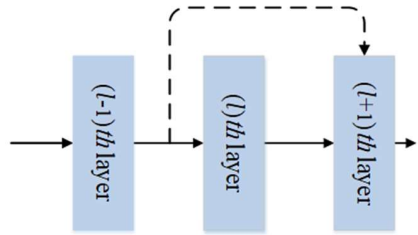

**Figure 1.** A building block of the residual network.

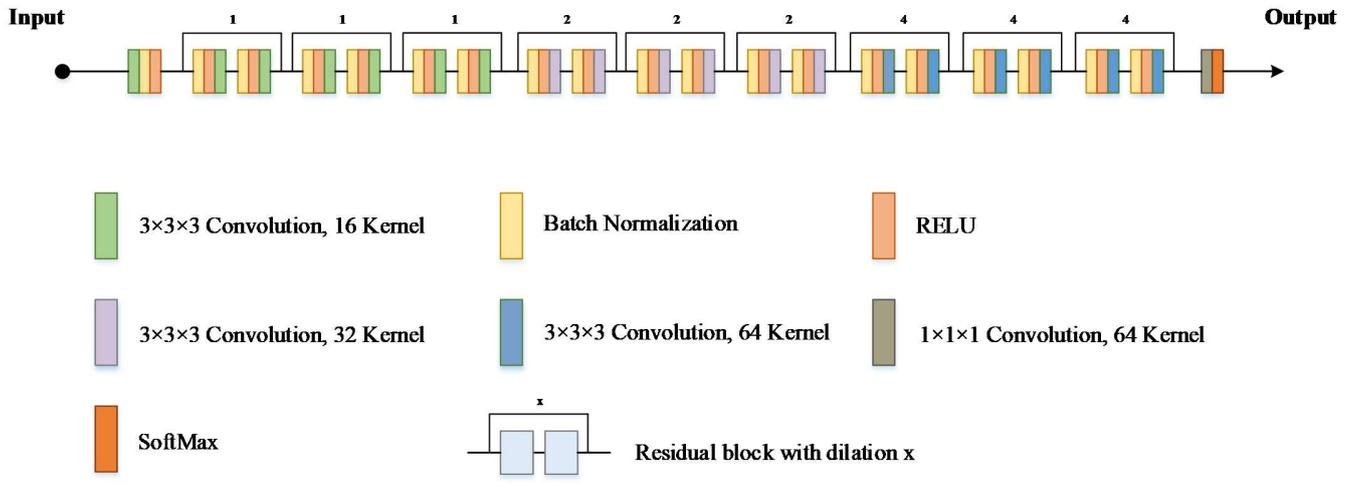

**Figure 2.** The architecture of ResNet model.

**Figure 3.** The architecture of the GAN model.

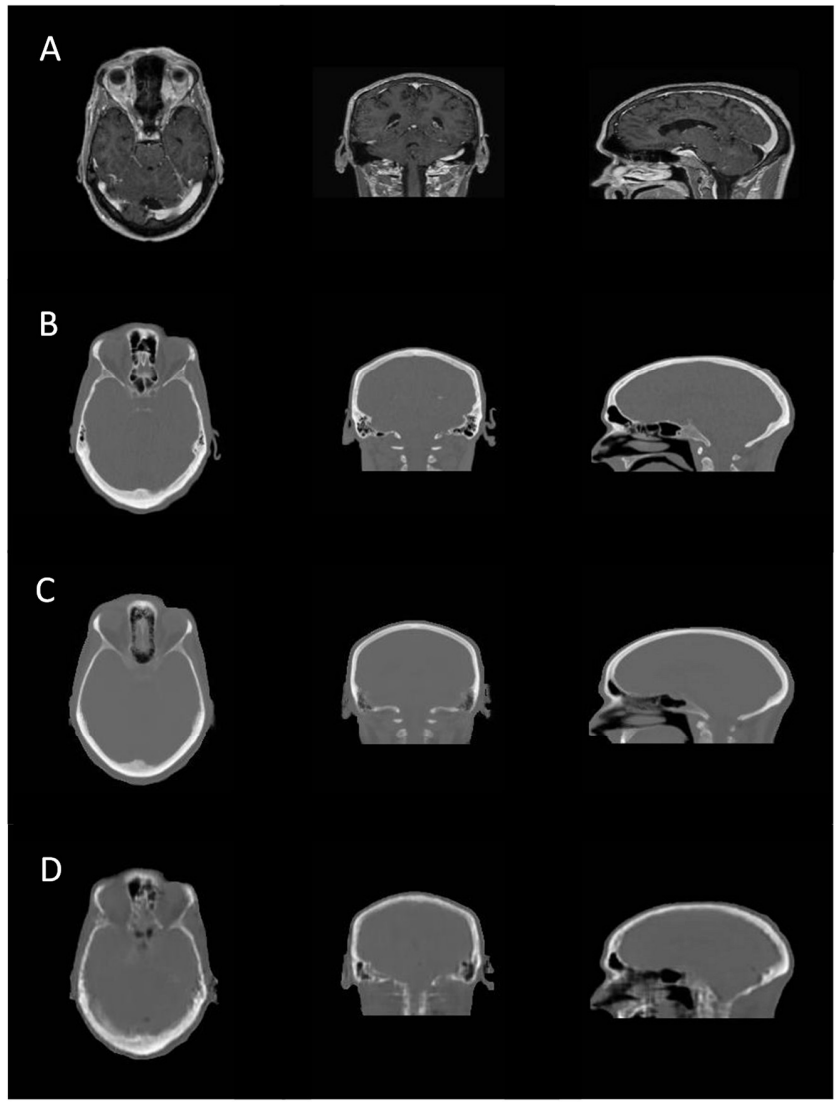

**Figure 4.** Qualitative comparison of sCT and reference CT images in three axial, sagittal, and coronal views: A) MR image, B) Reference CT, C) sCT generated by ResNet model D) sCT generated by GAN model.

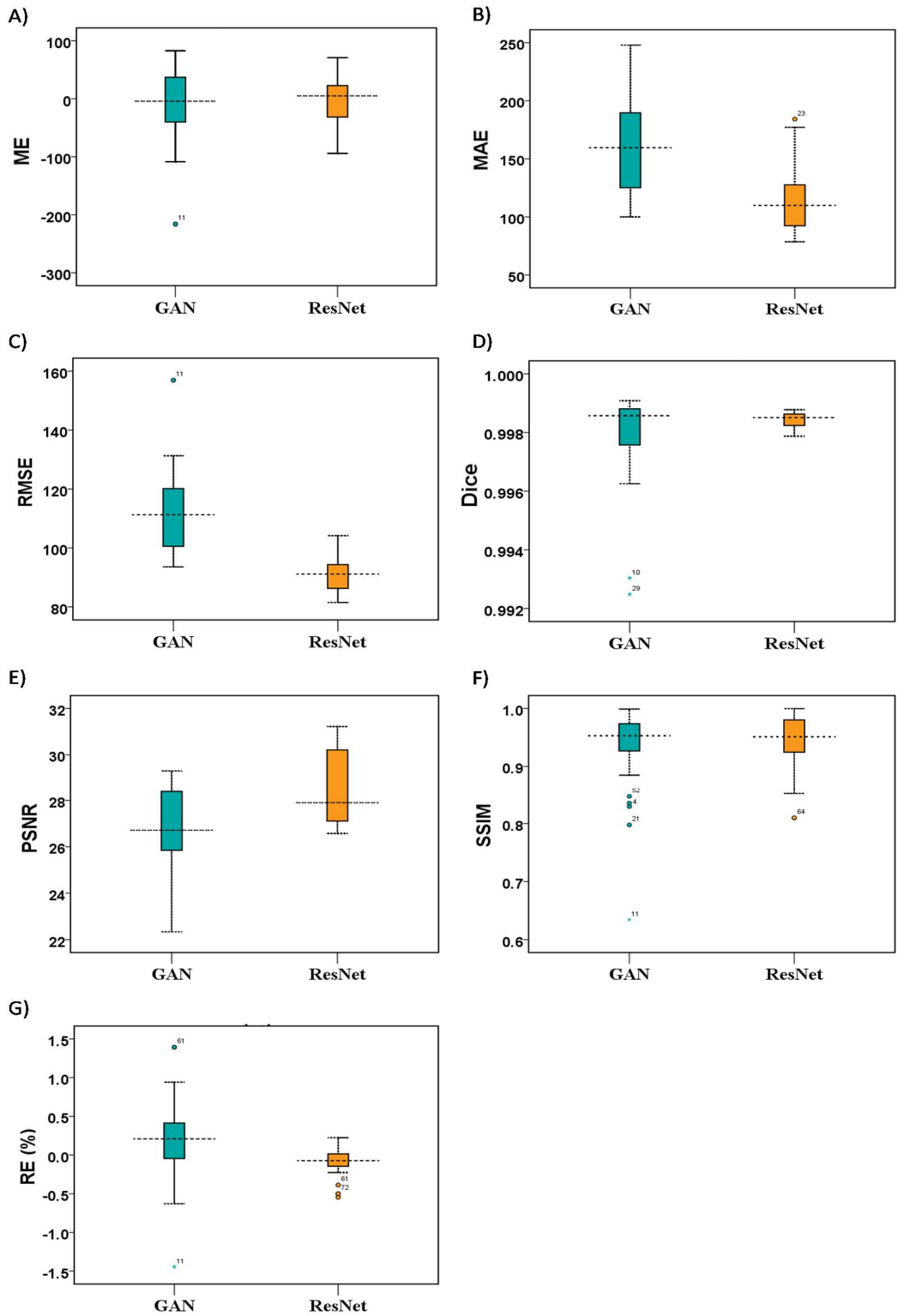

**Figure 5.** Boxplots of A) ME, B) MAE, C) RMSE, D) Dice, E) PSNR, F) SSIM and G) RE metrics between GAN and ResNet methods.

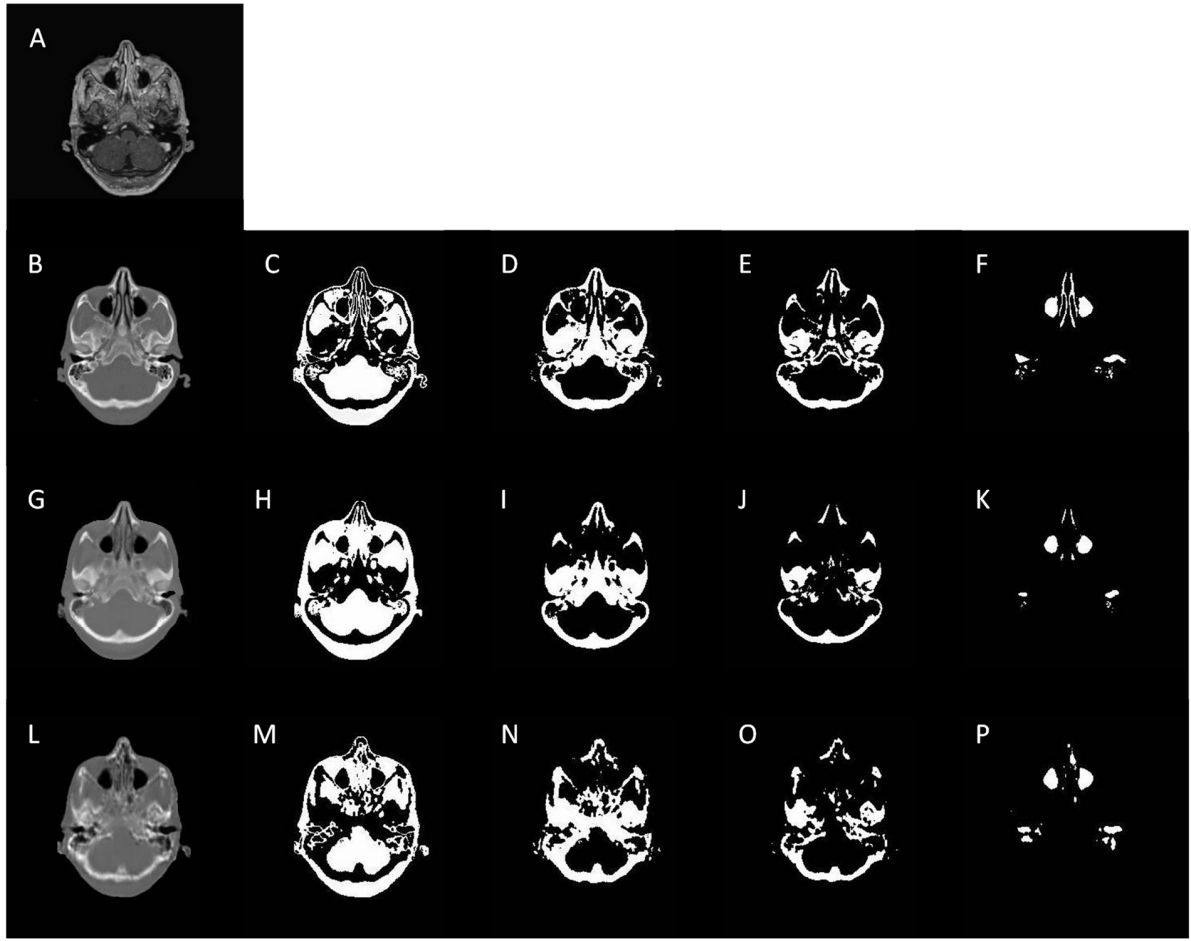

**Figure 6.** Representative slices of sCT generated by ResNet and GAN methods as well as ground truth CT images along with segmented soft tissue, total bone, cortical bone and air cavities A) MRI, B) Reference CT, C) Reference soft tissue mask, D) Reference total bone mask, E) Reference cortical bone mask, F) Reference air mask, G) ResNet CT, H) ResNet soft tissue mask, I) ResNet total bone mask, J) ResNet cortical bone mask, K) ResNet air mask, L) GAN CT, M) GAN soft tissue mask, N) GAN total bone mask, O) GAN cortical bone mak, P) GAN air mask

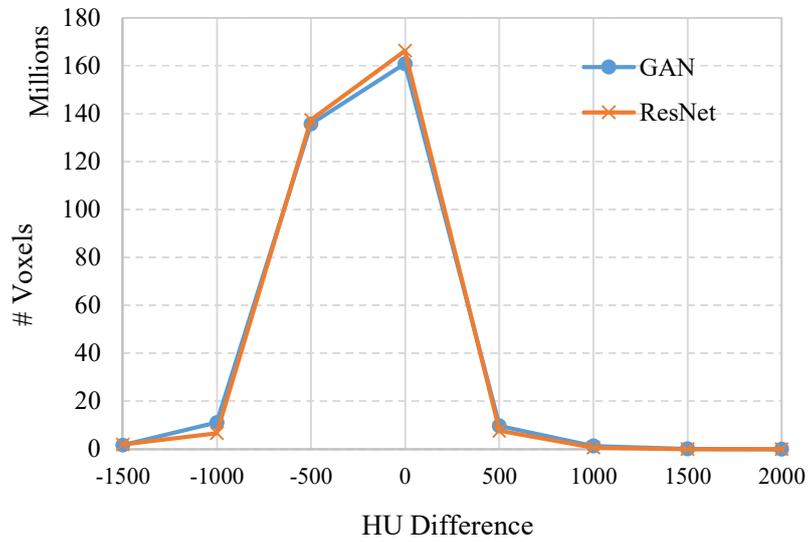

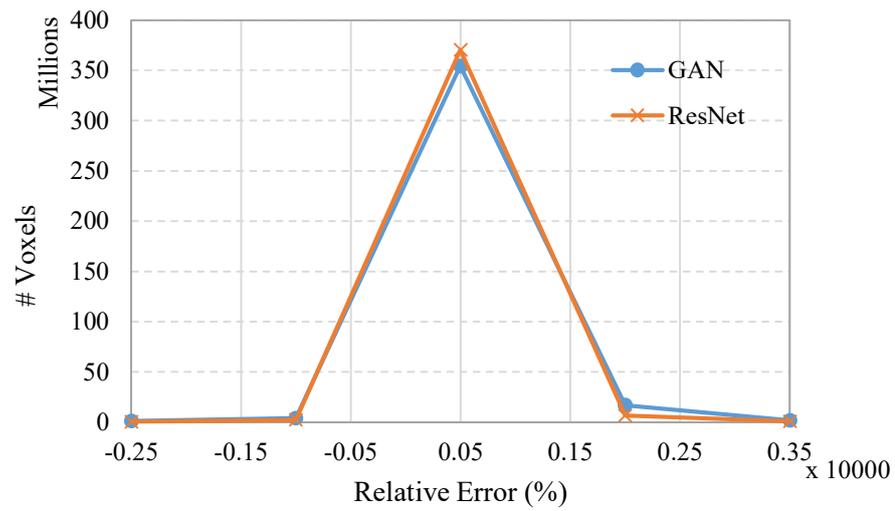

**Figure 7.** A) The difference of Hounsfield Unit, and B) the percentage of Relative Error between the reference CT and synthetic CT images resulted from GAN and ResNet methods.

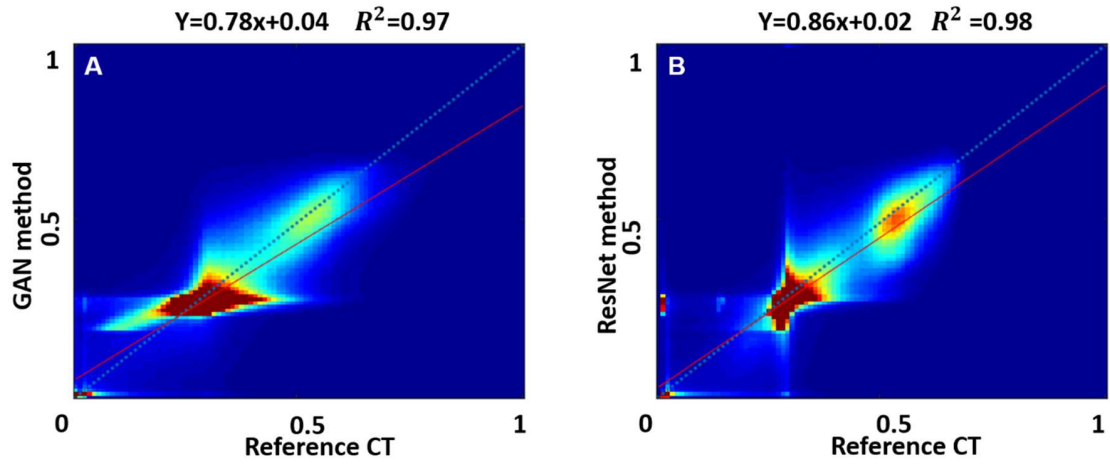

**Figure 8.** Joint histograms analysis of the sCT generated by the A) GAN & B) ResNet methods with respect to the reference CT over 86 subjects.